\documentclass[11pt]{article}

\usepackage[utf8]{inputenc}
\usepackage[T1]{fontenc}
\usepackage{lmodern}
\usepackage[a4paper,margin=1in]{geometry}
\usepackage{amsmath}
\usepackage{amssymb}
\usepackage{booktabs}
\usepackage{array}
\usepackage{graphicx}
\usepackage{enumitem}
\usepackage{url}
\usepackage[hidelinks]{hyperref}

\title{OpenAnt: LLM-Powered Vulnerability Discovery Through Code
Decomposition, Adversarial Verification, and Dynamic Testing}

\author{Nahum Korda \\ nahum.korda@knostic.ai \and Gadi Evron \\ gadi@knostic.ai}

\date{First pre-released: March 11, 2026 \\ Public pre-release: June 17, 2026}

\begin{document}

\maketitle

\begin{abstract}
Automated vulnerability discovery in large codebases remains challenging: traditional
static analysis produces high false-positive rates, while dynamic approaches such as fuzzing
require substantial infrastructure and often target narrow classes of bugs. Recent advances
in large language models (LLMs) enable semantic reasoning about program behavior, but
applying LLMs to repository-scale security analysis introduces challenges related to context
management, cost, and verification.

We present OpenAnt, an open-source vulnerability discovery system that integrates static
program analysis with LLM-based reasoning in a multi-stage pipeline. OpenAnt introduces
three key techniques. First, codebases are decomposed into self-contained analysis units
filtered by reachability from external entry points, reducing the analysis surface by up to
97\% while preserving attack-relevant code. Second, candidate vulnerabilities undergo adversarial verification through constrained attacker simulation, where the model evaluates
exploitability under realistic attacker capabilities. Third, findings are validated through dynamic verification, in which exploit environments are generated automatically, executed in
sandboxed containers, and discarded after use.

Evaluation on widely used open-source projects including OpenSSL, WordPress, and
Flowise shows that this architecture can identify previously unknown vulnerabilities while
maintaining manageable analysis cost and substantially reducing false positives. Our results
suggest that closed-loop vulnerability discovery pipelines, combining semantic reasoning with
exploit validation, provide a practical path toward scalable automated security analysis.
OpenAnt is released as open source under the Apache 2.0 license at \url{https://github.com/knostic/OpenAnt}.
\end{abstract}

\section{Introduction}

Static Application Security Testing (SAST) tools have been widely used for detecting vulnerabilities in source code for more than two decades. Modern SAST systems---including tools
such as Semgrep, CodeQL, Fortify, and Checkmarx---analyze program code to identify patterns
associated with security vulnerabilities, including unsafe data flows, unsanitized user inputs,
and misuse of security-sensitive APIs. Despite their widespread deployment in industrial development pipelines, prior empirical studies have identified several limitations that affect their
effectiveness in practice. In particular, large-scale evaluations of static analysis tools applied
to real-world codebases report substantial variability in detection precision across tools and
datasets. Empirical analyses of open-source software projects observe false-positive rates ranging from low single-digit percentages to over 40\%, depending on the analysis configuration,
benchmark suite, and evaluation methodology \cite{ref1,ref2}. Excessive warning noise reduces developer
trust in analysis results and has been identified as a key barrier to the effective adoption of
static analysis tools \cite{ref3,ref4}. Consistent with these findings, a recent large-scale survey of developers reports that only approximately 20\% of respondents actively use SAST tools in their
development workflow \cite{ref5}. When static analysis tools are used, developers frequently report
that large numbers of irrelevant warnings create alert fatigue and increase the likelihood that
important findings are ignored during vulnerability triage \cite{ref6}. Furthermore, rule-based static
analysis systems are inherently constrained by the expressiveness and coverage of their manually
engineered rules. Recent empirical work evaluating SAST tools for secure code review shows
that the majority of reported warnings may not correspond to the specific vulnerability under
investigation, limiting their usefulness for vulnerability identification and remediation \cite{ref7}.

These limitations motivate the exploration of alternative approaches to automated vulnerability detection. Recent advances in large language models (LLMs) have demonstrated strong
capabilities in code understanding and reasoning tasks. Modern code-oriented LLMs can analyze program syntax and semantics and perform a range of automated code analysis tasks \cite{ref8}.
Prior work further shows that LLM-based systems can infer aspects of developer intent from
surrounding code context and generate program logic consistent with natural language specifications \cite{ref9,ref10}. These capabilities suggest that LLM-based approaches may enable deeper
semantic reasoning about program behavior than traditional rule-based static analysis. However, existing research has largely evaluated LLM-based techniques on isolated code snippets,
synthetic benchmarks, or narrowly scoped analysis tasks. Their effectiveness for vulnerability
discovery in large, real-world codebases remains unclear.

Applying LLMs to repository-scale vulnerability analysis also introduces new technical challenges. Empirical studies of long-context reasoning show that model accuracy degrades when
relevant information appears in the middle of long inputs, a phenomenon commonly referred to
as the lost-in-the-middle effect \cite{ref11}. This limitation complicates the analysis of large software
repositories, even when models support extended context windows. In addition, LLM-based
analysis introduces computational costs that grow with the amount of analyzed code and may
produce incorrect conclusions when reasoning over incomplete program context. Addressing
these challenges remains an open problem for the application of LLMs to practical vulnerability
detection. In this work, we investigate how LLM-driven reasoning can be applied to vulnerability analysis while mitigating the limitations of both traditional static analysis and large-context
language models.

To address these challenges, we developed OpenAnt, a vulnerability analysis framework
designed to combine semantic reasoning with scalable repository-level analysis. OpenAnt organizes the analysis process as a multi-stage pipeline that progressively reduces the analysis
surface while increasing verification rigor for candidate vulnerabilities.

\subsection{Contributions}

This paper makes three primary contributions:

\textbf{Code decomposition into self-contained analysis units.} We present an automated pipeline
that decomposes source code into function-level analysis units, resolves cross-file dependencies, and constructs self-contained analysis contexts. The pipeline further restricts analysis
to functions reachable from externally exposed entry points. This decomposition significantly
reduces the analysis surface while preserving externally reachable---and therefore potentially
attackable---execution paths. In large repositories evaluated in this work, the approach reduces
the analyzed code volume by approximately 97\%.

\textbf{Adversarial verification through constrained attacker simulation.} We introduce a
verification methodology in which the language model evaluates candidate vulnerabilities by
simulating the capabilities of a remote attacker under realistic constraints (e.g., browser-only
access, no server-side privileges, and no administrative credentials). Under these constraints,
the model must reason about concrete exploitation paths and account for practical barriers such
as authentication mechanisms, input validation, and platform boundaries.

\textbf{Dynamic verification through ephemeral exploit generation.} We present a dynamic
verification mechanism that constructs vulnerability-specific exploit environments from first
principles, including attack code, execution infrastructure, and monitoring components. These
environments are executed in sandboxed Docker containers and discarded after execution, allowing the system to produce concrete evidence of exploitability without relying on pre-defined
exploit templates.

\section{System Architecture}

OpenAnt analyzes a source code repository through a six-stage pipeline that progressively narrows the analysis scope while increasing the depth of verification. Stages 1--2 perform static
program analysis without invoking language models. Stages 3--5 incorporate LLM-based reasoning, and Stage 6 performs runtime exploit validation.

\subsection{Stage 1: Code Parsing}

The first stage parses the repository and extracts all functions using language-specific abstract
syntax tree (AST) analysis. For each function, the parser records the function signature, function body, source file location, and call relationships. These relationships are used to construct
a bidirectional call graph linking callers and callees across files.

OpenAnt currently supports Python, JavaScript/TypeScript, Go, C/C++, Ruby, and PHP.
Python and Go are parsed using their native AST libraries. JavaScript and TypeScript are
parsed using ts-morph, while C/C++, Ruby, and PHP are parsed using tree-sitter grammars.
All parsers produce a unified intermediate representation used by subsequent pipeline stages.

\subsection{Stage 2: Unit Generation and Reachability Filtering}

The unit generation stage transforms extracted functions into self-contained analysis units,
which serve as the fundamental unit of security analysis in OpenAnt. Each unit contains:

\begin{itemize}
\item \textbf{Primary code:} the target function to be analyzed.
\item \textbf{Resolved dependencies:} functions called by the target function, resolved transitively
up to a configurable depth (default: three levels). Dependencies are inlined from across
the repository to produce a self-contained code block that can be analyzed without loading
the full codebase.
\item \textbf{Entry-point metadata:} information indicating whether the function is reachable from
externally supplied input, such as HTTP handlers, command-line argument processors,
WebSocket handlers, or file-reading interfaces.
\end{itemize}

\textbf{Reachability filtering.} Not all code in a repository is reachable from external input. Internal
utilities, helper functions used exclusively by tests, and administrative tools are typically not
part of the attack surface for a remote adversary. OpenAnt therefore identifies entry points---
functions that accept externally supplied input---through a combination of decorator analysis
(e.g., \texttt{@app.route}, \texttt{@router.get}, \texttt{@Get()}), parameter pattern recognition (e.g., \texttt{request.args},
\texttt{req.query}), and framework-specific conventions.

Starting from identified entry points, OpenAnt performs a breadth-first traversal of the call
graph to mark all reachable functions. Only functions reachable from these entry points are
retained as analysis units for subsequent LLM-based stages.

This filtering step provides the primary cost reduction mechanism in the pipeline. For
example, in a repository containing 15,232 extracted functions (OpenSSL), reachability filtering
reduces the analysis set to 390 units (a 97\% reduction). In Grafana (18,500 functions), the
filtering stage produces 994 reachable units, corresponding to a 94.6\% reduction.

\subsection{Stage 3: Exposure Classification}

Each reachable unit is analyzed by an LLM agent (Claude Sonnet) that iteratively explores
the codebase using tool-assisted navigation. The agent may search for callers, read function
implementations, and trace call paths in order to classify the unit's security exposure.

Each unit is assigned one of four categories:

\begin{center}
\begin{tabular}{ll}
\toprule
\textbf{Classification} & \textbf{Description} \\
\midrule
Exploitable & Vulnerable code reachable from external input \\
Vulnerable-internal & Potentially unsafe code not reachable from user input \\
Security control & Defensive logic such as validation, sanitization, or authentication \\
Neutral & Code with no apparent security relevance \\
\bottomrule
\end{tabular}
\end{center}

Only units classified as exploitable proceed to vulnerability detection.

In the OpenSSL repository example, this stage reduces the 390 reachable units to 49 externally exposed units, representing an additional 87\% reduction.

\subsection{Stage 4: Vulnerability Detection}

Each exposed unit is analyzed by an LLM (Claude Opus) using a language-agnostic prompt.
The prompt instructs the model to answer three questions:

\begin{enumerate}
\item What does this code do?
\item Where does the input originate?
\item What security risk may arise from this behavior?
\end{enumerate}

The model produces a structured assessment containing one of five possible findings.

\begin{center}
\begin{tabular}{ll}
\toprule
\textbf{Finding} & \textbf{Meaning} \\
\midrule
vulnerable & Exploitable without effective protection \\
bypassable & Protection exists but may be circumvented \\
inconclusive & Security posture cannot be determined \\
protected & Potentially dangerous operations guarded by effective controls \\
safe & No security-sensitive operations identified \\
\bottomrule
\end{tabular}
\end{center}

A key design choice is that the detection prompt is language-agnostic. The same prompt is
used across Python, JavaScript/TypeScript, Go, C/C++, Ruby, and PHP. Security reasoning
at the semantic level---tracing externally controlled input to sensitive operations---generalizes
across programming languages, allowing a single prompt to support multiple language ecosystems.

\subsection{Stage 5: Adversarial Verification}

All units that receive any finding in Stage 4---including those classified as safe or protected---are
submitted to adversarial verification. This stage evaluates whether the identified conditions can
be exploited by a realistic attacker.

\textbf{Attacker simulation.} The language model is instructed to simulate a remote attacker operating under realistic constraints:

\begin{itemize}
\item browser-based interaction only
\item no server-side access
\item no administrative credentials
\item no ability to modify server files
\end{itemize}

Under these constraints, the model attempts to construct exploitation paths for the analyzed
code. During this process the model may search the codebase, inspect function implementations,
and trace call relationships. Each attempted attack is documented step-by-step, including
whether each step is feasible and what conditions prevent exploitation.

\textbf{Multi-approach exploration.} Initial experiments showed that single-path reasoning frequently missed exploitable conditions. A function may resist one attack strategy while remaining vulnerable to another. The verification prompt therefore requires the model to attempt
multiple exploitation approaches before concluding that a vulnerability is absent.

\textbf{Victim requirement.} To avoid false positives based on self-impact scenarios, the verification
process requires that a valid vulnerability must produce harm affecting a party other than the
attacker. Scenarios in which the attacker can only affect their own data are therefore excluded.

The output of this stage is a structured document describing the analyzed function, each
exploitation attempt, whether the attempt succeeded or failed, and the final assessment.

\subsection{Stage 6: Dynamic Verification}

Findings that remain plausible after adversarial verification are submitted to dynamic testing
through automated exploit generation.

For each candidate vulnerability, an LLM (Claude Sonnet) generates:

\begin{itemize}
\item a Dockerfile specifying the execution environment and dependencies
\item a test script that attempts to reproduce the exploit
\item a requirements file specifying required packages
\item optionally, a Docker Compose configuration for multi-service scenarios such as SSRF testing
\end{itemize}

The generated exploit environment is executed inside an isolated Docker container with the
following restrictions:

\begin{itemize}
\item read-only filesystem
\item 512 MB memory limit
\item single CPU allocation
\item no privilege escalation
\item execution timeout of 120 seconds
\end{itemize}

The container must output a single JSON object to standard output classifying the result
as: CONFIRMED, NOT\_REPRODUCED, BLOCKED, INCONCLUSIVE, or ERROR.

If execution results in an ERROR (e.g., build failure or runtime crash), the error message is
returned to the LLM together with the original finding and generated test code. The model
then generates a corrected test environment, and execution is retried for up to three iterations.

\textbf{Ephemeral execution.} All generated artifacts---including test scripts, Docker images, and
intermediate files---are discarded after execution. Each exploit environment is constructed
specifically for the analyzed vulnerability and removed after producing a result. This design
ensures that the system produces concrete evidence of exploitability without maintaining a
persistent library of exploit templates.

\section{Evaluation Methodology}

\subsection{The Benchmark Contamination Problem}

Standard security benchmarks---including the Juliet Test Suite \cite{ref12}, the OWASP Benchmark \cite{ref13},
and curated vulnerability datasets---have served as the primary evaluation methodology for
static analysis tools for more than a decade. However, for LLM-based vulnerability detection
systems, these benchmarks introduce significant methodological concerns.

The training corpora of large language models contain vast quantities of publicly available source code, technical documentation, and security research. Benchmarks such as Juliet,
OWASP Benchmark, and many publicly reproduced CVE examples are openly available and
therefore likely to appear in the training data of modern code-oriented LLMs \cite{riddell}. Under these conditions, high benchmark accuracy may reflect memorization rather than genuine vulnerability
reasoning. This concern has been demonstrated empirically: studies of LLM benchmark leakage
show that models can achieve dramatically higher performance on contaminated benchmark
samples compared to unseen data. For example, recent analysis across software engineering
benchmarks reports pass rates up to 4.9$\times$ higher on leaked samples compared to non-leaked
ones \cite{ref14}. Broader surveys of contamination in LLM evaluation emphasize that benchmark
validity requires explicit contamination analysis when models may have been exposed to the
benchmark during training \cite{ref15,ref16}.

The reliability of existing vulnerability datasets is further complicated by data quality issues.
An empirical study of commonly used vulnerability datasets found that 20--71\% of vulnerability
labels were inaccurate and 17--99\% of samples were duplicated, raising concerns about their suitability for evaluating machine learning models \cite{ref17}. Subsequent work introducing the PrimeVul
dataset, which focuses on realistic vulnerabilities extracted from real repositories, reports that
LLM-based detectors perform substantially worse on realistic vulnerability detection tasks than
on synthetic benchmarks. In particular, detection accuracy on synthetic datasets such as Juliet
was observed to be 10.5\% higher on average than on real-world datasets \cite{ref18,ref19}.

Additional characteristics of synthetic benchmarks further limit their suitability for evaluating LLM-based systems. For example, the Juliet Test Suite contains over 80,000 test cases with
systematic naming conventions (e.g., \texttt{CWE89\_SQL\_Injection\_\_*}) that reveal the vulnerability
type directly in the filename. Similar patterns exist in OWASP Benchmark test cases. These
artifacts enable models to infer the vulnerability category without analyzing the code itself,
making the benchmarks vulnerable to shortcut learning.

Because benchmark datasets may both appear in model training data and contain structural
artifacts, evaluation based solely on benchmark performance cannot reliably measure a model's
vulnerability reasoning capability.

\subsection{Evaluation Through Real-World Vulnerability Discovery}

These limitations motivate an alternative evaluation strategy based on real-world vulnerability
discovery. Instead of measuring performance on curated benchmark datasets, the evaluation
focuses on the ability of the system to identify previously unknown vulnerabilities in actively
maintained software projects.

This approach evaluates vulnerability detection systems under realistic conditions in which
the analyzed code was not constructed for benchmarking and may not appear in the training
data of modern language models. Successful discoveries therefore provide direct evidence that
the system can identify security flaws in complex, evolving codebases rather than recognizing
patterns present in synthetic datasets.

Real-world vulnerability discovery also evaluates multiple aspects of a security analysis system simultaneously, including code comprehension, vulnerability reasoning, and the ability to
distinguish exploitable flaws from benign patterns. Unlike synthetic benchmarks, which typically encode a single vulnerability in a controlled context, real-world software often contains
complex interactions between components, partial mitigations, and defensive logic that complicate vulnerability detection.

For these reasons, the evaluation presented in this paper focuses on OpenAnt's ability to
identify vulnerabilities in real-world open-source projects.

\subsection{Scope of Detection}

OpenAnt is designed to detect vulnerabilities arising from unsafe propagation of externally
controlled input to security-sensitive operations, including injection attacks, path traversal,
server-side request forgery (SSRF), authorization bypass, and cross-site scripting (XSS). These
vulnerabilities share a common structural pattern in which attacker-controlled input reaches a
dangerous sink without adequate validation.

The detection prompt therefore focuses on tracing the origin of input and its use within
the program, enabling the model to reason about exploitability in terms of realistic attacker
capabilities.

Many vulnerabilities reported in CVE datasets do not follow this pattern. Instead, they arise
from missing functionality, specification violations, or complex cross-component logic errors.
Examples include missing resource limits (CWE-770), absent cleanup operations (CWE-772),
protocol parsing inconsistencies such as HTTP request smuggling, or vulnerabilities that emerge
only through interactions across multiple subsystems.

These issues require reasoning about expected program behavior or protocol specifications,
rather than identifying unsafe data flows. As a result, recall measurements based on curated vulnerability datasets---many of which contain substantial numbers of logic bugs and specification
violations---do not accurately reflect the detection scope of OpenAnt.

Accordingly, the evaluation emphasizes real-world vulnerability discovery, where the system's ability to identify exploitable input-driven vulnerabilities can be validated through adversarial verification and dynamic exploit reproduction.

\section{Evaluation}

We evaluate OpenAnt on eight real-world open-source projects spanning six programming languages. The evaluation measures four properties of the system:

\begin{enumerate}
\item Scalability --- the ability to analyze large repositories through progressive filtering.
\item Detection capability --- the ability to identify potentially exploitable security flaws.
\item False-positive mitigation --- the effectiveness of adversarial verification in eliminating
impractical findings.
\item Practical exploitability --- the rate at which findings can be confirmed through automated dynamic testing.
\end{enumerate}

All experiments execute the full six-stage pipeline described in Section 2 without manual
intervention.

\subsection{Evaluation Setup}

\textbf{Models.} As described in Section 2, Stages 3 and 6 use Claude Sonnet 4 (exposure classification and dynamic exploit generation), and Stages 4 and 5 use Claude Opus 4 (vulnerability detection and adversarial verification).

Temperature is set to 0 for all stages to maximize determinism. No system-level caching is
used; each analysis unit is processed independently.

\textbf{Project Selection.} Projects were selected according to three criteria:

\begin{enumerate}
\item Active maintenance and adoption (all projects exceed 10,000 GitHub stars).
\item Diversity across programming languages and application architectures.
\item Presence of externally reachable attack surface, such as HTTP endpoints, API handlers,
or public interfaces.
\end{enumerate}

\begin{center}
\begin{tabular}{lll}
\toprule
\textbf{Project} & \textbf{Language} & \textbf{Application Type} \\
\midrule
OpenSSL & C & Cryptographic library \\
Flowise & JavaScript & LLM workflow platform \\
eShopOnWeb & PHP & Web application \\
n8n & TypeScript & Workflow automation \\
WordPress & PHP & Content management system \\
object-browser & Go & Web application \\
paperless-ngx & Python & Document management system \\
Rails & Ruby & Web framework \\
\bottomrule
\end{tabular}
\end{center}

The evaluation set includes web applications, automation platforms, a cryptographic library,
and a mature web framework. This diversity exposes the pipeline to multiple vulnerability
classes, including injection flaws, authorization bypasses, filesystem access issues, and memory
safety errors.

\textbf{Terminology.} Throughout this section we use the following terminology:

\begin{center}
\begin{tabular}{ll}
\toprule
\textbf{Term} & \textbf{Definition} \\
\midrule
Flagged & Unit identified as potentially vulnerable in Stage 4 \\
Confirmed & Finding surviving adversarial verification (Stage 5) \\
Dynamically verified & Finding reproduced through automated exploit generation (Stage 6) \\
\bottomrule
\end{tabular}
\end{center}

\subsection{Key Evaluation Results}

Across eight real-world repositories comprising 64,132 functions, OpenAnt progressively reduced
the analysis surface to a small set of security-relevant candidates and produced reproducible
exploit evidence for a substantial fraction of findings.

The evaluation demonstrates four key properties of the system.

\textbf{Scalable repository analysis.} Static reachability filtering reduced the initial analysis set
from 64,132 functions to 2,281 reachable units (96.4\% reduction) before any LLM reasoning was
invoked. Subsequent exposure classification further reduced the set to 586 externally exploitable
units, representing less than 1\% of the original codebase.

\textbf{Practical vulnerability discovery.} Adversarial verification identified 190 vulnerability candidates across the evaluated repositories. These findings span more than 30 vulnerability classes,
including authorization bypasses, server-side request forgery (SSRF), injection vulnerabilities,
and path traversal flaws.

\textbf{Exploit reproducibility.} Automated dynamic testing successfully reproduced 144 vulnerabilities (75.8\%) through generated exploit environments executed in sandboxed containers.
These exploits were produced automatically without handcrafted test cases.

\textbf{Cost efficiency through staged filtering.} The full evaluation across all repositories cost
\$1,461.25, with exposure classification accounting for 72.9\% of total cost due to the agentic
code exploration required to trace call relationships and security exposure. Without reachability filtering, the same analysis would have cost approximately \$23,700, demonstrating the
importance of the pipeline's static filtering stages.

More broadly, these results highlight a key distinction between OpenAnt and most existing vulnerability analysis tools. Traditional SAST systems---and many recent LLM-based
analyzers---primarily perform static vulnerability detection, reporting patterns that may correspond to security flaws. In contrast, OpenAnt implements a closed-loop vulnerability discovery
pipeline in which detection is followed by adversarial exploit reasoning and automated exploit
generation. This design enables the system to move beyond identifying suspicious code patterns
toward producing concrete exploit evidence, significantly reducing false positives and providing
actionable security findings.

\subsection{Pipeline Funnel}

The following table presents the progressive filtering across all eight projects.

\begin{center}
\resizebox{\textwidth}{!}{%
\begin{tabular}{llrrrrrr}
\toprule
\textbf{Project} & \textbf{Language} & \textbf{Functions} & \textbf{Reachable} & \textbf{Exploitable} & \textbf{Flagged (S4)} & \textbf{Confirmed (S5)} & \textbf{Dynamic (S6)} \\
\midrule
OpenSSL & C & 15,232 & 390 & 49 & 28 & 3 & 1 \\
Flowise & JavaScript & 3,564 & 303 & 160 & 129 & 85 & 64 \\
eShopOnWeb & PHP & 999 & 23 & 7 & 4 & 3 & 3 \\
n8n & TypeScript & 16,009 & 794 & 196 & 118 & 62 & 49 \\
WordPress & PHP & 12,177 & 393 & 93 & 67 & 20 & 13 \\
object-browser & Go & 1,268 & 110 & 16 & 9 & 4 & 4 \\
paperless-ngx & Python & 1,065 & 179 & 47 & 19 & 11 & 9 \\
Rails & Ruby & 13,818 & 89 & 18 & 2 & 2 & 1 \\
\midrule
Total & & 64,132 & 2,281 & 586 & 376 & 190 & 144 \\
\bottomrule
\end{tabular}%
}
\end{center}

Across all repositories, reachability filtering eliminates 83--99\% of functions, dramatically
reducing the amount of code analyzed by LLM stages.

For example, in OpenSSL:

Only 0.02\% of original functions ultimately produced confirmed findings.

\subsection{Vulnerability Discovery}

The pipeline identified 190 vulnerability candidates that remained exploitable after adversarial
verification (Stage 5). Of these, 144 vulnerabilities were independently reproduced through
automated dynamic exploit generation (Stage 6). The remaining findings represent plausible
exploitation paths that could not be dynamically reproduced within the constraints of automatically generated test environments, typically due to complex runtime dependencies or
environmental requirements.

\begin{center}
\begin{tabular}{lrr}
\toprule
\textbf{Stage} & \textbf{Units Remaining} & \textbf{Reduction} \\
\midrule
Parsing & 15,232 & --- \\
Reachability filtering & 390 & 97.4\% \\
Exposure classification & 49 & 99.7\% \\
Vulnerability detection & 28 & 99.8\% \\
Adversarial verification & 3 & 99.98\% \\
\bottomrule
\end{tabular}
\end{center}

These findings span more than 30 vulnerability categories.

\begin{center}
\resizebox{\textwidth}{!}{%
\begin{tabular}{lrrl}
\toprule
\textbf{Vulnerability Type} & \textbf{Confirmed} & \textbf{Dyn. Confirmed} & \textbf{Projects} \\
\midrule
IDOR / Missing Authorization & 29 & 21 & Flowise, paperless-ngx \\
Mass Assignment / Privilege Escalation & 26 & 20 & Flowise, n8n \\
SSRF & 25 & 18 & Flowise, n8n, WordPress \\
Path Traversal / File Read / Zip Slip & 18 & 16 & Multiple \\
XSS / Content Injection & 17 & 10 & WordPress, n8n \\
SQL / NoSQL Injection & 11 & 8 & Flowise, n8n \\
Unauthenticated Endpoint / Auth Bypass & 6 & 5 & n8n \\
Resource Exhaustion / DoS & 6 & 2 & WordPress \\
Header Injection & 6 & 5 & Flowise \\
Other (22 types) & 33 & 28 & Various \\
\bottomrule
\end{tabular}%
}
\end{center}

The diversity of findings across six languages demonstrates that semantic vulnerability reasoning generalizes across programming ecosystems.

Importantly, the detection prompt was not modified between languages during evaluation.
The same prompt template was applied to all repositories regardless of programming language,
indicating that vulnerability reasoning based on externally controlled input, data flow, and attacker capabilities generalizes across programming ecosystems rather than relying on language-specific prompt engineering.

Workflow automation platforms (Flowise, n8n) exhibited the highest vulnerability density,
reflecting the complexity of systems that allow users to execute arbitrary workflows and interact
with external services.

In contrast, mature frameworks such as Rails and cryptographic libraries such as OpenSSL
produced relatively few confirmed findings.

\subsection{Dynamic Verification}

All 190 findings confirmed by adversarial verification were submitted to automated dynamic
testing.

\begin{center}
\begin{tabular}{lrr}
\toprule
\textbf{Outcome} & \textbf{Count} & \textbf{Percentage} \\
\midrule
CONFIRMED & 144 & 75.8\% \\
INCONCLUSIVE & 24 & 12.6\% \\
NOT\_REPRODUCED & 13 & 6.8\% \\
ERROR & 8 & 4.2\% \\
BLOCKED & 1 & 0.5\% \\
\bottomrule
\end{tabular}
\end{center}

The 75.8\% dynamic confirmation rate provides independent validation of the adversarial
verification stage.

Vulnerability classes requiring direct input manipulation show the highest confirmation
rates:

\begin{center}
\begin{tabular}{lr}
\toprule
\textbf{Vulnerability} & \textbf{Dynamic Rate} \\
\midrule
Command Injection & 100\% \\
Path Traversal & 88.9\% \\
Authentication Bypass & 83.3\% \\
Mass Assignment & 76.9\% \\
\bottomrule
\end{tabular}
\end{center}

In contrast, timing-dependent issues such as race conditions are difficult to reproduce automatically.

\subsection{False Positive Mitigation}

Adversarial verification is the primary mechanism for eliminating theoretical vulnerabilities that
cannot be exploited in practice.

Across all projects:

\begin{itemize}
\item 376 findings flagged in Stage 4
\item 190 confirmed after attacker simulation
\item 49.5\% elimination rate
\end{itemize}

Common rejection causes include:

\begin{enumerate}
\item Input sanitization preventing attacker-controlled data flow.
\item Authentication barriers blocking exploitation.
\item Vulnerabilities affecting only the attacker's own data (violating the victim requirement).
\item Platform protections such as browser same-origin policy or cloud storage access controls.
\end{enumerate}

Dynamic testing provides a lower bound on verification accuracy: 144 of 190 confirmed
findings produced functional exploits.

\subsection{Cost Analysis}

The total evaluation cost across all eight projects was \$1,461.25.

\begin{center}
\begin{tabular}{lr}
\toprule
\textbf{Project} & \textbf{Total Cost} \\
\midrule
OpenSSL & \$442.66 \\
Flowise & \$229.09 \\
eShopOnWeb & \$6.37 \\
n8n & \$392.84 \\
WordPress & \$243.84 \\
object-browser & \$31.35 \\
paperless-ngx & \$91.86 \\
Rails & \$23.25 \\
\bottomrule
\end{tabular}
\end{center}

Exposure classification (Stage 3) accounts for 72.9\% of total cost, reflecting the agentic
exploration required to understand call relationships and security exposure.

Without reachability filtering, analyzing all extracted functions at the median Stage 3 cost
would have cost approximately \$23,700. Static filtering therefore reduces total analysis cost by
more than 96\%, making repository-scale LLM-based security analysis economically feasible.

\section{Related Work}

Automated vulnerability discovery has been studied extensively across several research traditions, including pattern-based static analysis, dynamic testing through fuzzing, and more
recently LLM-assisted security analysis. OpenAnt builds on insights from these areas while
combining them in a unified pipeline that integrates static program analysis, semantic reasoning, adversarial verification, and dynamic exploit generation.

\textbf{Pattern-Based Static Application Security Testing.} Traditional static application security testing (SAST) tools detect vulnerabilities through syntactic pattern matching and data-flow analysis. Systems such as Semgrep and CodeQL rely on manually authored rules that
encode vulnerability patterns and propagation paths \cite{ref20,ref21}. These tools can analyze large
codebases efficiently and integrate well into development pipelines. Large-scale evaluations, including the NIST Static Analysis Tool Exposition (SATE), document substantial variability in
both detection capability and false positive rates across tools and configurations \cite{ref1,ref22}.

However, rule-based approaches are constrained by the expressiveness and completeness of
their rule sets. Vulnerabilities that do not match predefined patterns may remain undetected,
while benign code may trigger alerts when patterns match without sufficient contextual information \cite{ref6,ref7}. OpenAnt differs from these approaches in two important ways. First, vulnerability identification is based on semantic reasoning about attacker-controlled input and program
behavior, rather than rule matching. Second, candidate findings are subjected to explicit exploitability verification, reducing false positives by requiring plausible attacker execution paths.

\textbf{LLM-Assisted Code Security Analysis.} Recent research has explored the use of large
language models for automated security analysis. Several systems leverage LLMs to reason
about code semantics and identify potential vulnerabilities beyond the scope of pattern-based
detectors \cite{ref18,ref19}. Google's ``Big Sleep'' project demonstrated that LLM-assisted reasoning
can identify subtle memory safety bugs that evade traditional testing methods \cite{ref23}. Industrial systems such as Anthropic's Claude Code Security and OpenAI's Aardvark apply similar
approaches to large-scale repository analysis \cite{ref24,ref25}.

These systems show that LLMs can reason about software security at a semantic level, but
they typically rely on static reasoning alone and therefore inherit a key challenge of traditional
SAST: distinguishing theoretical vulnerabilities from practically exploitable ones. OpenAnt
extends this line of work by introducing a closed-loop verification architecture in which vulnerability hypotheses generated by an LLM are subsequently evaluated through adversarial
attacker simulation and automated exploit generation. This design shifts the analysis objective
from identifying suspicious patterns to producing concrete exploit evidence.

\textbf{LLM-Based Autonomous Penetration Testing.} Another line of work investigates the
use of LLM agents to autonomously exploit vulnerabilities. Recent studies show that LLM-based agents can execute penetration testing tasks such as blind SQL injection, authentication
bypass, and privilege escalation when provided with iterative feedback from target systems \cite{ref26,ref27,ref28,ref29}. Systems such as PentestGPT demonstrate that structured prompting and tool-assisted reasoning can significantly improve the ability of language models to perform multi-step
security tasks \cite{ref29}. Other work has shown that LLM agents can autonomously exploit known
vulnerabilities or attack web applications under constrained interaction models \cite{ref26,ref27}.

OpenAnt incorporates ideas from this line of work in its adversarial verification stage, where
the model simulates a realistic attacker attempting to exploit candidate vulnerabilities under defined constraints. Unlike penetration testing agents, however, OpenAnt performs this reasoning
directly on source code, enabling vulnerability analysis before deployment.

\textbf{Fuzzing and Dynamic Vulnerability Discovery.} Dynamic analysis techniques such as
fuzzing discover vulnerabilities by executing programs with generated inputs. Modern fuzzing
systems including AFL, libFuzzer, and OSS-Fuzz have successfully uncovered large numbers of
memory safety vulnerabilities in widely deployed software \cite{ref30,ref31}.

Fuzzing excels at identifying bugs that manifest through observable runtime behavior but
typically requires compilation, instrumentation, and execution infrastructure specific to each
target \cite{ref31}. Furthermore, fuzzing effectiveness depends heavily on input generation strategies
and coverage feedback mechanisms \cite{ref30,ref31}. OpenAnt's dynamic verification stage serves a different purpose. Rather than exploring large input spaces through random mutation, the system
generates targeted exploit scenarios derived from earlier semantic analysis. This approach allows dynamic validation across multiple programming languages without requiring specialized
instrumentation.

\textbf{Automated Exploit Generation.} Automated exploit generation has been studied extensively in the context of cyber-autonomy systems \cite{mayhem,angr}. The DARPA Cyber Grand Challenge demonstrated fully automated pipelines capable of discovering vulnerabilities and generating exploits
in binary programs \cite{ref32}. More recent work explores the use of language models to assist exploit construction by reasoning about vulnerability conditions and generating proof-of-concept
payloads \cite{ref26,ref27}.

OpenAnt adopts a similar philosophy but applies it at the source code analysis stage. Exploit
environments are generated automatically and executed within sandboxed containers to confirm
or refute candidate vulnerabilities. Importantly, generated artifacts are ephemeral and exist only
long enough to produce a verification result.

\section{Limitations}

Although OpenAnt demonstrates promising results for automated vulnerability discovery, several limitations remain.

\textbf{Dependence on LLM Capabilities.} The detection and verification stages rely on the
reasoning capabilities of the underlying language model. As a result, OpenAnt inherits the
strengths and weaknesses of contemporary LLMs \cite{ref8,ref11,ref18,ref19}. Certain vulnerability classes
remain difficult for the system to detect reliably, particularly those involving subtle semantic
conditions or domain-specific invariants. More broadly, prior work on LLM-based vulnerability detection also reports uneven performance across vulnerability categories and substantial
sensitivity to evaluation setting and benchmark design \cite{ref18,ref19}.

Future work may address these limitations through specialized prompts, domain-specific
training data, or hybrid approaches that combine symbolic analysis with LLM reasoning.

\textbf{Cost and Scalability.} Although the multi-stage filtering pipeline substantially reduces analysis cost, large repositories still require significant computational resources. In the evaluation
presented in this paper, the largest repository analyses cost several hundred dollars each due
primarily to the agentic exploration required during exposure classification. This limitation reflects a broader challenge in repository-scale LLM analysis: long-context reasoning and iterative
tool use remain computationally expensive \cite{ref11}.

Organizations with large numbers of repositories may therefore need to prioritize targets
or perform incremental analysis. Improvements in model efficiency and caching strategies may
reduce these costs over time.

\textbf{Dynamic Verification Coverage.} The dynamic verification stage confirms exploitability
by generating and executing proof-of-concept exploits. While effective for many vulnerability
classes, this approach has limitations. Certain vulnerabilities require complex runtime environments, external infrastructure, or precise timing conditions that are difficult to reproduce
automatically in isolated containers. For example, race conditions and distributed system exploits may require coordinated interactions across multiple services that exceed the scope of
automatically generated test environments. Similar constraints are well known in dynamic
analysis and exploit-generation systems more broadly \cite{ref30,ref31,ref32}.

As a result, some valid vulnerabilities may remain plausible but dynamically unconfirmed.

\textbf{Absence of Formal Guarantees.} OpenAnt provides empirical evidence of exploitability
rather than formal proofs of vulnerability or safety. Adversarial verification reduces false positives by requiring realistic attack scenarios, and dynamic testing provides concrete confirmation
when exploits succeed. However, the absence of successful exploitation does not guarantee that
a vulnerability is impossible. This limitation is shared by both LLM-based security analysis
and dynamic testing systems more broadly \cite{ref18,ref19,ref30,ref31}.

Formal verification methods could complement this approach by providing stronger guarantees for critical components.

\textbf{Context Window Constraints.} Despite the use of code decomposition and dependency
resolution, some analysis units may exceed the context window limits of the underlying model.
In such cases, relevant code may be truncated or require iterative retrieval. While the system
attempts to mitigate this limitation through dependency pruning and tool-assisted exploration,
context limitations remain a practical constraint for very large functions or deeply nested call
chains. This issue is consistent with prior findings on long-context degradation in language
models \cite{ref11}.

Future work may explore hierarchical analysis techniques or specialized models optimized
for long-context program reasoning.

\section{Conclusion}

Automated vulnerability discovery remains a challenging problem: traditional static analysis
produces large numbers of false positives, while dynamic approaches such as fuzzing require
significant infrastructure and often target narrow classes of vulnerabilities. Recent advances in
large language models offer new opportunities for semantic program reasoning, but applying
LLMs to repository-scale security analysis introduces challenges related to context management,
cost, and verification.

This paper presented OpenAnt, an LLM-powered vulnerability discovery system that combines static code decomposition, adversarial attacker simulation, and automated exploit generation in a unified pipeline. By decomposing repositories into reachable analysis units, OpenAnt
reduces the analysis surface dramatically before applying LLM reasoning. Adversarial verification then evaluates candidate vulnerabilities from the perspective of a constrained attacker,
while dynamic verification produces executable exploit environments to confirm exploitability.

Evaluation on real-world open-source projects demonstrates that this architecture can identify previously unknown vulnerabilities while maintaining a manageable analysis cost and substantially reducing false positives. More broadly, the adversarial verification methodology introduced in this work provides a practical mechanism for distinguishing theoretical vulnerabilities
from exploitable ones, a challenge that affects many LLM-based security analysis systems.

OpenAnt suggests that closed-loop vulnerability discovery pipelines---combining semantic
reasoning with exploit validation---may provide a viable path toward scalable automated security analysis. Future work may extend this approach to additional vulnerability classes, integrate
richer program analysis techniques, and explore applications beyond vulnerability detection.

OpenAnt is released as open source under the Apache 2.0 license at \url{https://github.com/knostic/OpenAnt}.

\end{document}